\documentclass{article}
\usepackage{amsmath,amssymb,amsfonts}
\usepackage{amsbsy}
\usepackage{enumerate}
\usepackage{graphicx}
\usepackage{tikz}
\usepackage{datetime}

\newcommand{\be}{\begin{equation}}
\newcommand{\ee}{\end{equation}}

\def\bea{\begin{eqnarray}}
\def\eea{\end{eqnarray}}
\def\bean{\begin{eqnarray*}}
\def\eean{\end{eqnarray*}}

\newcommand{\barr}{\begin{array}}
\newcommand{\earr}{\end{array}}

\newcommand{\bed}{\begin{displaymath}}
\newcommand{\eed}{\end{displaymath}}
\newcommand{\bal}{\begin{array}{ll}}
\newcommand{\eal}{\end{array}}

\def\mc#1{\mathcal#1}

\begin{document}

\centerline{\Large Tri-Bi-Maximal Mixing in Asymmetric Textures}
\vskip .2cm
\centerline{Pierre Ramond}
\centerline{University of Florida}
\vskip .3cm

\centerline{ \large In the Honor of A. Balachandran's eighty-fifth Birthday}


\section{Syracuse to Trieste}
In honor of Professor Ayalam Balachandran who had the patience to teach me theoretical physics.  

When accepted to graduate school at Syracuse University, I wanted to study General Relativity with Peter Bergmann.  After an arduous first year in transition from engineering to physics, I took a wonderful course from Professor MacFarlane's  on Advanced Quantum Mechanics  which turned me into  a budding Particle Physicist.  George Sudarshan, my first adviser, asked me this  simple question: ``Does the Spin-Statistic Theorem apply to infinite component wave equations?", and left shortly after for a sabbatical in India. This is how I got acquainted with Majorana's equation and Streater and Whightman's ``PCT, Spin and Statistics, and All That". Professor Sudarshan came back to India and told me that Ivan Todorov had built a counter example. So much for that! 
\vskip .3cm
\noindent 
Left adrift,  I was rescued by Professor Balachandran who, I found out,  lived in momentum space. He proposed a problem: given a four-point S-matrix for massive particles, is it possible to continue into the Mandelstam Triangle and emerge to a different channel. This was a concrete task, where I learned of the Appel-Kamp\'e de Feriet's  orthogonal polynomials in the triangle, but the physics problem was much too difficult because of the cuts, and I was left with an incomplete answer, much to my dismay. During this process, Bal  continued his encouragements, and when away, with helpful,  lengthy and detailed letters. With Bal, my fellow graduate student Bill Meggs, and Jean Nuyts, I participated in three published papers. This was almost not enough: after four years, I was on the job market, but was not picked until much later when Bob Wilson decided at the last moment to initiate a theory group at the newly approved National Accelerator Laboratory (NAL), later known as FermiLab. 
\vskip .3cm
\noindent 
 I will be forever grateful to Bal for securing me a summer appointment at Trieste's  International Center for Theoretical Physics that Professor Salam had founded. There, I had the good fortune to meet my collaborator Jean Nuyts and Hirotaka Sugawara, who introduced me to the amazing mathematical structure behind  the Veneziano model. Hirotaka and I worked hard to extract  the triple Reggeon vertex using tensor methods, only to find  Sciuto's elegant version using ladder operators - thankfully we never published our version, but I was hooked on Veneziano's Dual Resonance model.
 
 \vskip .3cm
 \noindent
  I am honored to contribute to this volume in Bal's honor. Without him, I would have gone back to engineering, and perhaps 
  been wealthier, but surely not as fulfilled nor as happy!
  
  \section{Asymmetric Texture}
The presence of two large angles in the neutrino mixings is, I believe,  the most important  yet most neglected puzzle of particle physics. Whenever Nature displays a breakdown in previously accepted patterns, there lies the path to further progress. Quarks and leptons, with similar gauge couplings, show totally different mixings. Quarks follow small mixing angles while neutrinos display only one small mixing angle and two large mixing angle whose values approximate ``geometric" values. This led to an inspired approximation\cite{TBM}  of the lepton mixing matrix, 

$$\mc U^{}_{PMNS}=\mc U^{(-1)}_{}\mc U^{}_{TBM},\qquad
 \mc U^{}_{\rm TBM}=\begin{pmatrix}\sqrt{\dfrac{2}{3}}&\dfrac{1}{\sqrt{3}}&0\\[0.8em] -\dfrac{1}{\sqrt{6}}&\dfrac{1}{\sqrt{3}}&\dfrac{1}{\sqrt{2}}\\[1em] 
\dfrac{1}{\sqrt{6}}&-\dfrac{1}{\sqrt{3}}&\dfrac{1}{\sqrt{2}}\end{pmatrix},
$$ 
where $\mc U^{(-1)}_{}$ comes from the diagonalization of the charged lepton Yukawa matrix. In the $SU_5$ Grand-Unified extension of the Standard Model, $\mc U^{(-1)}_{}$  should generate  a ``Cabibbo Haze" that fully account for the third (reactor) angle. In $SU_5$, is   summarized by the elegant Georgi-Jarlskog\cite{GJ} construction,
\be
\mathcal U^{(-1)}_{}=\mathcal U_\mathrm{CKM}(c~\rightarrow~ -3c),
\ee
where $c$ is the (22) element of the down quarks Yukawa matrix, and the GUT-Scale relations 
$$
~m_\tau=m_b,\,\, m_\mu=3m_s,\,\, m_e=\frac{1}{3}m_d, ~~\longrightarrow~~   
\det Y^{(-1/3)}=\det Y^{(-1)}$$
Before the reactor angle had been measured, a plethora of models\cite{review} satisfying these relations appeared in the literature, with predictions for the reactor angle around  one-third the Cabibbo angle. When measured\cite{reactor} to be twice  this value, the TBM approximation was (too) quickly forgotten.
\vskip .5cm
\noindent 
As a SuperString person, I always look for mathematical beauty in physics, following my hero Paul Dirac\cite{PAM};  the  TBM matrix is beautiful since it makes one think of angles between crystal faces, finite group theory, etc ... .  

\vskip .5cm
\noindent Closer examination reveals that these models assume that all Yukawa matrices are symmetric, an assumption based not on theory but on convenience. The Georgi-Jarlskog relation of Eq(1) no longer applies if the charged lepton and down quarks Yukawa matrices are  {\it asymmetric}.
\vskip .5cm 
\noindent 
The search for asymmetric Yukawa matrices that reproduce the same G-J relations while violating Eq(1) was conducted\cite{RRX} in the ``Damn The Torpedoes" mode, an expression often used by Feynman (after Farragut). It meant do the calculation and do not worry - if difficulties arise, fix them later if possible. The Yukawa  matrices turn out to be fine-tuned and models will be required to explain their fine-tunings.
\vskip .5cm 
\noindent Such matrices  exist\cite{RRX} but in a highly constrained form,

$$
Y^{(2/3)}_{}=\begin{pmatrix}\lambda^8&0&0\cr 0&\lambda^4&0\cr 0&0&1\end{pmatrix},\,\,
Y^{(-1/3)}_{}=\begin{pmatrix}{bd\lambda^4}&a\lambda^3&{b\lambda^3}\cr a\lambda^3&{c\lambda^2}&g\lambda^2\cr 
{d\lambda}&g\lambda^2&1\end{pmatrix},\,\, 
Y^{(-1)}_{}=\begin{pmatrix}{bd\lambda^4}&a\lambda^3&{d\lambda}\cr a\lambda^3&{-3c\lambda^2}&g\lambda^2\cr {b\lambda^3}&g\lambda^2&1\end{pmatrix}
$$
where $a=c=1/3,b=.306,g=0.811, d=2a/g$ are Wolfenstein-like parameters. 

Our ``Asymmetric Texture", with TBM hypothesis yields, 
$$\theta^{}_{12}=39.81^\circ,\quad 6.16^\circ \text{ above PDG};\qquad \theta^{}_{23}= 42.67^\circ,\quad 2.90^\circ \text {below PDG},$$
and $\theta_{13}$ overshoots its PDG value:
$$ \sin\theta_{13}=\sin\theta_{\rm reactor}= 0.184,\quad \text{  compared to} =\sin\theta^{PDG}_{\rm reactor}=0.145
$$
It is fortunate that an overshoot can be corrected by adding to TBM a CP-violating phase 
 
 $$\mathcal U^{}_\mathrm{TBM}~\longrightarrow ~U^{}_\mathrm{TBM}(\delta)=
{\rm Diag} (1,1,e^{i\delta})\,\mathcal U^{}_\mathrm{TBM},$$
with

  $$
\sin|\theta^{}_\mathrm {reactor}|~\rightarrow~\sin|\theta^{}_\mathrm {reactor}(\delta)|=\dfrac{\lambda}{3\sqrt 2}\left| 1+\dfrac{2e^{i\delta}}{g}\right|<\sin|\theta^{}_\mathrm {reactor}|.
$$
The phase is thus determined by setting  $\theta_{\rm reactor}$ at its PDG value, up to a sign, 
\vskip .5cm 
 $ \delta^{}_{CP}=\pm 1.32\pi,~~~{\rm compared ~with ~ global~ fits:  }~~~~ \delta^{PDG}_{CP}= 1.36
 \begin{matrix} \scriptstyle +0.20\cr \scriptstyle -0.16\end{matrix}\,\pi.$
\vskip .3cm
\noindent
It is satisfying that this one  phase corrects both solar and atmospheric mixing angles to their experimentally-consistent values, 
 \vskip .3cm

 $\theta^{}_{\rm Atm}=44.9^\circ ~(0.66^\circ~ \text{below PDG}),\quad \theta^{}_{\rm Solar}=34.16^\circ~ (0.51^\circ \text{above PDG}).$

 \newpage
 \section {Family Symmetry}
As mentioned above the asymmetric texture was designed to satisfy the G-J $SU_5$ relations, and the CKM matrix for quark mixings and the Gatto relation. With a phased TBM it predicts the magnitude of the phase. It also contains many fine-tuned patterns and values. To alleviate these problems,  a discrete family symmetry is required. The general features of this model\cite{PRRSX} are:
\vskip .5cm

 \noindent - Standard Model Chiral Fields  transform  as $SU_5~{\rm irreps}:~{\bf \bar 5},\,{\bf 10},\,{\bf 1}$ 
 \vskip .2cm
 
\noindent -  No Tree Level Electroweak Yukawa Couplings. 
\vskip .2cm
\noindent - Vector-Like Fermion Messengers coupling to SM fields and ``Familons",
\vskip .2cm
\noindent -  Familons: scalar SM singlets with only family charges
\vskip .2cm
\noindent -  Non-Abelian Family Symmetry with CG coefficients  to single out asymmetry.
\vskip 1cm

\noindent The Family Symmetry of choice is the Non-Abelian ``Frobenius" discrete group, the semi-direct product  $\mc Z_3$, $T^{}_{13}=(Z_{13}\rtimes Z_3)$, with  two inequivalent  triplet representations  
\vskip .2cm
$ Z_{13}\rtimes Z_3~~ {\rm Irreps}:~~{\bf 3_1},\,{\bf 3_2},\,{\bf 1'},\,{\bf \bar 3_1},\,{\bf \bar 3_2},\,{\bf \bar 1'},\,{\bf 1}$
\vskip .2cm
\noindent It arises from continuous $G_2$ via the simple group $PSL(2,13)$:
\vskip .2cm
$ Z_{13}\rtimes Z_3=T^{}_{13}\subset PSL(2,13)\subset G^{}_2 $
\vskip .2cm
\noindent - The chiral matter is assigned to different family triplets:  $({\bf \bar 5},{\bf 3_1})+({\bf 10},{\bf 3_2})$
\vskip .2cm
\noindent - Electroweak-breaking Higgs fields: $  H\sim ({\bf \bar 5},{\bf 1}),\,\,H'\sim ({\bf \overline { 45}},{\bf 1})$\vskip .2cm
\noindent- $ \text{ Vectorlike Messengers:} $
\vskip .3cm
$\Delta\sim ({\bf 5},{\bf 3_2});\, \text{ explains charged lepton (down-quarks) Yukawa with  H} $
\vskip .2cm
$\Sigma\sim ({\bf 10},{\bf 3_1})~\text{ generates the G-J term with H'} $
\vskip .2cm
$\Gamma\sim ({\bf\overline {10}},{\bf 3_2}),\,\Theta\sim({\bf \overline {10}},{\bf \bar 3_1})~\text { generate up-quarks Yukawas}$
\vskip .3cm

\noindent - $ \text { Familons: four} ~\sim ({\bf 1},{\bf 3_2}),\quad {\rm two} ~\sim ({\bf 1},{\bf 3_2})
$
\vskip .2cm

\noindent - $\text{ Familon vacuum values lie along simple directions}:$
\vskip .2cm
\hskip.7cm $ (1,0,0), (0,1,0),(0,0,1),(0,1,1,),(1,0,1)$
\vskip .5cm
\noindent
The effect of the model is to explain the fine-tunings in terms of simple direction in the familons' vacuum manifold.

\newpage
\noindent The next step is to generate the Seesaw sector. To wit, as in the original Seesaw,  one  upgrades from $SU_5$ to $SO_{10}$ where the right-handed (sterile) neutrinos appear:
\vskip .3cm

$SO_{10}\supset SU_5\times U(1):~~~~~ {\bf 16}={\bf\bar 5}+{\bf 10}+{\bf 1};~~{\bf 10}_v={\bf\bar 5}+{\bf 5}.$
\vskip .3cm
\noindent The chiral assignment  upgrades naturally to
\vskip .3cm
$SU_5\times T_{13}:~~({\bf 10},{\bf 3_2})+({\bf \bar 5},{\bf 3_1})~~\longrightarrow~~SO_{10}\times T_{13}: ~~~({\bf 16},{\bf 3_2})+({\bf 10}_v,{\bf 3_1})$
\vskip .3cm
\noindent It reproduces the original chiral fields once the conjugate quintet in the spinor ${\bf 16}$ couples to the quintet in the vector ${\bf 10}_v$   generate a vector-like mass through a familon interaction, and three sterile neutrinos $\overline N$ which transform as $({\bf 1},{\bf 3_2})$ under $SU_5\times T_{13}$. 
\vskip .3cm
\noindent 
Assume  a dimension-four coupling of the sterile neutrinos to one familon:  $\overline N\,\overline N\varphi_{\mc M}$. $T_{13}$ invariance  allows $\varphi_{\mc M}\sim ({\bf 1},{\bf 3_2}).$ 
\vskip .3cm
\noindent 
In the vacuum direction, $<\varphi_{\mathcal M}>_0=M(1,-1,1)$, the  Majorana mass of the three sterile neutrinos is diagonalized by the TBM matrix:
\vskip .3cm
$$ 
\mc U_{TBM}^{t}\frac{1}{\mc M}\mc U_{TBM}^{}=\frac{1}{M}
{\rm Diag}( 1,-\frac{1}{2},1)$$
\vskip .3cm
\noindent  With $T_{13}$  a simple vacuum structure yields TBM diagonalization.

\vskip .3cm
\noindent The Seesaw Mechanism\cite{seesaw} requires  the  diagonalization of

$$
\mc S=\mc  D\,\frac{1}{\mathcal M}\,\mc  D^t,
$$ 
where  $\mc D$ is the Dirac mass stemming from the coupling of $\overline  F\sim ({\bf\bar 5},{3_1})$ with $\overline N$: $\overline F\,\overline N\varphi_{\mc D}$, with $\varphi _{\mc D}\sim {\bf\bar 3_1}.$

\vskip .3cm
\noindent The simple familon vacuum: $<\varphi_{\mc D}>=a(1,-1,1)$, $\mc S$  is diagonalized by TBM,

$$
{\mathcal S}=\frac{a^2}{M}{\mathcal U_{\rm TBM}^{}}{\rm Diag}( 1,-\frac{1}{2},1){\mathcal U_{\rm TBM}^t},
$$
\vskip .5cm
\noindent from which we extract the light neutrino masses: $m_{\nu_3}=2m_{\nu_2}$ implies normal ordering, but the degeneracy $m_{\nu_1}=m_{\nu_3}$ implies no $(\nu_1-\nu_3)$ oscillations, {\it in obvious contradiction with experiment!}
\newpage
\noindent Could there be more sterile neutrinos? The sequence from $SU_5~\rightarrow~SO_{10}$  continues  to the exceptional group $E_6$ as we showed long ago.   

\vskip .5cm
\noindent 
Fourth Sterile Neutrino: $SO_{10} \rightarrow E_6$ and the group theory follows,

\vskip .2cm
\noindent
- $E_6\supset SO_{10}\times U_1:  ~~~{\bf 27}= {\bf 16}+{\bf 10}+{\bf 1}$
\vskip .2cm
\noindent
- $T_{!3}$'s mother group $PSL(2,13)$ contains a complex septet:~~~${\bf 7}={\bf 3_1}+{\bf 3_2}+{\bf 1'}$
\vskip .3cm
\noindent A natural marriage of $E_6$ and $PSL(2,13)$ ensues with chiral assignments 
\vskip .3cm

$SO_{10}\times T_{13}: ~~({\bf 16},{\bf 3_2})+({\bf 10},{\bf 3_1})~\longrightarrow~E_6\times T_{13}:~~({\bf 16},{\bf 3_2})+({\bf 10},{\bf 3_1})+({\bf 1},{\bf 1'})$

\vskip .3cm
\noindent With a fourth sterile neutrino, the Seesaw matrix contains a new addition
\vskip .3cm
 
$\mc S~\rightarrow~\mc S'=\mc S+\mc W\mc W^t$,
\vskip .3cm
\noindent where $\mc W\mc W^t$ has two zero eigenvalues. Of the three choices, only one agrees with  oscillations, with required vacuum value $<\mc W^t>_0\propto (2,\, e^{i\pi},\, e^{i\delta}).$ 
\vskip .3cm
\noindent Using oscillations, this model predicts normal ordering of the light neutrino masses:
\vskip .5cm
$$m_{\nu_1}=27.6~ {\rm meV},~~m_{\nu_2}=28.9~{\rm meV},~~m_{\nu_3}=57.8~{\rm meV}$$
\vskip 1cm

\noindent We found recently \cite{asym2} a version of the asymmetric texture, with the smaller family symmetry $T_7=\mc Z_7\rtimes \mc Z_3$.  Its ancestry  originates in continuous $G_2\supset PSL(2,7)\supset T_7$. 
\vskip .2cm
\noindent $PSL(2,7)$ contains a real septet  irrep ${\bf 7}={\bf 3}+{\bf \bar 3}+{\bf 1}$, which suggests a similar mixture based on $E_6$ and $PSL(2,7)$:
\vskip .2cm
\noindent  $ E_6\times T_{7}:~~~~~~({\bf 16},{\bf 3})+({\bf 10},{\bf \bar 3})+({\bf 1},{\bf 1})$
 These curious chiral assignments whose modular symmetries point to  $G_2$ may construct a path to coset manifolds in eleven dimensions.

\section{Acknowledgements}
I wish to thank Bal again for allowing me to spend my life chasing  the fabrics of our Universe. 
I thank Professors P\'erez and Stuart,  as well as Drs Rahat and Xu  for helpful discussions. This research was supported in part by the Department of Energy under Grant No. DE-SC0010296.

\end{document}